\documentclass[runningheads]{llncs}

\usepackage[hidelinks]{hyperref}

\usepackage{comment}
\usepackage{graphicx}

\newcommand{\specialcell}[2][c]{%
  \begin{tabular}[#1]{@{}l@{}}#2\end{tabular}}

\newcommand{\CC}{C\nolinebreak\hspace{-.05em}\raisebox{.4ex}{\tiny\bf +}\nolinebreak\hspace{-.10em}\raisebox{.4ex}{\tiny\bf +}}

\begin{document}

\title{Lessons learned in a decade of \\ research software engineering GPU applications}

\titlerunning{Lessons learned in RSEing GPU applications}

\author{Ben van Werkhoven\orcidID{0000-0002-7508-3272}\inst{1} \and Willem Jan Palenstijn\orcidID{0000-0003-0511-4763}\inst{2} \and Alessio~Sclocco\orcidID{0000-0003-3278-0518}\inst{1}}
\institute{Netherlands eScience Center, Amsterdam, the Netherlands \\
\email{b.vanwerkhoven@esciencecenter.nl}, \email{a.sclocco@esciencecenter.nl} \and
Centrum Wiskunde \& Informatica (CWI), Amsterdam, the Netherlands \\
\email{w.j.palenstijn@cwi.nl}}


\maketitle

\begin{abstract}

After years of using Graphics Processing Units (GPUs) to accelerate scientific applications in fields as varied as
tomography, computer vision, climate modeling, digital forensics, geospatial databases, particle physics, radio astronomy, and 
localization microscopy, we noticed a number of technical, socio-technical, and non-technical challenges that
Research Software Engineers (RSEs) may run into. 
While some of these challenges, such as managing different programming languages within a project, or having to deal
with different memory spaces, are common to all software projects involving GPUs, others are more typical of 
scientific software projects.
Among these challenges we include changing resolutions or scales, maintaining an application
over time and making it sustainable, and evaluating both the obtained results and the achieved performance.
%

\end{abstract}


\section{Introduction}


Since the introduction of programmable units in Graphics Processing Units (GPUs) scientists have been using GPUs because of their raw compute power and high 
energy efficiency~\cite{zwart2007high}. Today, many of the top500 supercomputers are equipped with GPUs~\cite{top500} and GPUs are a driving force behind the 
recent surge in machine learning~\cite{chetlur2014cudnn,abadi2016tensorflow,chen2015mxnet}. However, developing GPU applications requires computations to be 
parallelized using specialized programming languages, and to achieve high performance requires to understand the underlying hardware~\cite{ryoo2008optimization}. 
As such, many GPU applications have been developed by Research Software Engineers (RSEs)~\cite{baxter2012research} that have specialized in this field.
%
This paper presents an overview of our experiences and of the lessons learned from developing GPU applications for scientific research in a wide range 
of domains, with different computational requirements, and a variety of programming languages.



Developing GPU applications requires to make software architectural choices that will be costly to change once implemented, concerning
device memory management, host and device code integration across programming languages, and synchronizing and maintaining multiple versions 
of computational kernels.
This is due to the fact that the commonly used GPU programming systems do not allow developers to easily explore the whole design space.
%
In addition, GPU kernels exhibit large design spaces with different ways to map computations to threads and thread blocks, different layouts to use in 
specialized memories, specific hardware features to exploit, and values to select for thread block dimensions, work per thread, and loop 
unrolling factors. As such, auto-tuning is often necessary to achieve optimal and portable performance.

This paper also presents challenges and lessons learned specific to GPU {\em research software}.
%
Research software is defined by Hettrick et al.~\cite{hettrick2014uk} as software that is used to generate, process or analyze results that are intended to 
appear in scientific publications. Research software is often developed using short-lived grant-based research funding, and 
therefore its development is focused on new features instead of reliability and maintainability~\cite{goble2014better}. Finally, several surveys have shown
that research software is in large 
part developed by scientists who lack a formal education or even interest in software engineering best practices~\cite{hannay2009scientists,hettrick2014uk}.
The experiences we share in this paper are based on both short-lived collaborations between RSEs and scientists and on collaborations where an RSE was embedded 
in a research group for a longer period of time.

Advancements in hardware, compilers, and testing frameworks present new opportunities to apply software engineering best practices to GPU 
applications. However, it is important to realize that GPU research software is often based on existing software developed by scientists 
with no formal training in software engineering.
%
The software sustainability of GPU research software remains an open challenge as GPU programming remains a specialized field and
RSEs are often only involved during short-lived collaborative projects.

The reason to move code to the GPU is often to target larger, more complex problems, which may require the development of new methods to
operate at higher resolutions or unprecedented problem scales.
%
Evaluating the results of these applications often requires carefully constructed test cases and expert knowledge from the original developers.
%
Designing fair and reproducible performance experiments that involve applications using different languages, compilers, and hardware is a difficult task. This 
has led to the publication of controversial performance comparisons, and currently hinders RSEs in publishing about their work.

The rest of this paper is organized as follows. Section~\ref{sec:applications} lists the applications that we use as case studies in this paper.
Section~\ref{sec:gpuprogramming} presents the challenges and lessons learned that apply to GPU programming in any context, whereas 
Section~\ref{sec:gpurseing} focuses on those specific to a research software engineering context. Finally,
Section~\ref{sec:conclusions} summarizes our conclusions.

\section{Case Studies}\label{sec:applications}

In this section, we introduce some of the scientific GPU applications that we developed, which we use as case studies throughout the paper.
These applications span a wide range of scientific domains and host programming languages, as well as target hardware platforms, ranging
from GPU-enabled supercomputers to workstations and desktop computers equipped with one or more consumer grade GPUs.
The remainder of this section contains a brief description of each application, and highlights some of the challenges encountered; a summary
of the characteristics of said applications is provided in Table~\ref{tab:applications}.

{\bf 2D single molecule localization microscopy (SMLM)}~\cite{microscopy} is a MATLAB application, implementing a 
template-free particle fusion algorithm based on an all-to-all registration, which provides robustness against individual 
mis-registrations and underlabeling. The method combines many different observations into a single super-resolution reconstruction, 
working directly with localizations rather than pixelated images.
The application is very compute intensive as it uses several quadratic algorithms for computing registration scores on the GPU.

{\bf 3D Geospatial Data Explorer (GDE)}~\cite{pointinpolygon} is a spatial database management system which provides in-situ data 
access, spatial operations, and interactive data visualization for large, dense LiDAR data sets. The system was designed to handle a 
LiDAR scan of the Netherlands of 640 billion points, combined with cadastral information. The main GPU kernel implements a 
point-in-polygon algorithm for selecting data points within a selected shape. To optimize the latency of this operation, 
host to device transfers are overlapped with execution on the GPU, and auto-tuning is used extensively.

{\bf AMBER}~\cite{amber}, the Apertif Monitor for Bursts Encountered in Real-time, is a fully auto-tuned radio-astronomical pipeline for detecting 
Fast Radio Bursts and other single pulse transients, developed in \CC{} and OpenCL\@.
All the main computational components of AMBER are accelerated and run on the GPU, and the pipeline is deployed and used in production at the Westerbork
radio telescope~\cite{verheijen2008}.
The main technical challenge developing this application has been processing, in real-time, the large amount of data produced by the telescope,
while the main non-technical challenge has been to properly validate the application's results.

{\bf ASTRA Toolbox}~\cite{astra2011gpu} is a toolbox of high-performance GPU primitives for 2D and 3D tomography aimed at researchers and algorithm developers.
The basic forward and backward projection operations are GPU-accelerated, and callable via a \CC-interface from MATLAB and Python to enable 
building new algorithms. The main challenges were the trade-off between flexibility and performance, making effective use of memory caching,
and the initial lack of reference code.

{\bf Common image source identification (CISI)}~\cite{nfi} is a digital forensics tool that clusters a collection of digital photos 
based on whether they were acquired using the same image sensor. The GPU application implements a pipeline that extracts image 
noise patterns, as well as two algorithms for computing similarity scores. This application was developed in collaboration with the Netherlands 
Forensics Institute, where most of the software development happens in Java,
as such the host code for the GPU application is written in Java as well.

{\bf KM3NeT L0-Trigger} is a GPU pipeline designed to process unfiltered data from the KM3NeT neutrino telescope~\cite{jong2010}. Most of the 
existing processing happens on filtered, so called L1 data. The main issue in the design and implementation of this pipeline was the 
fact that the algorithms that operate on the L1 data were not suitable for processing L0 data. As such, we had to design a 
new pipeline that correlates L0 hits, clusters them, and classifies whether a neutrino event has occurred.

\begin{table}[t!]
    \centering
    \scriptsize
    \begin{tabular}{|l|l|l|l|l|l|}
\hline
application name	&	scientific domain	&	language	&	main bottleneck	&	target hardware	&	existing GPU code \\
\hline
\hline
2D \& 3D SMLM	&	microscopy	&	MATLAB	&	compute	&	desktop/server	&	No \\
\hline
3D GDE	&	\specialcell{geospatial \\ databases}	&	\CC	&	latency	&	server	&	No \\
\hline
AMBER	&	radio astronomy	&	\CC	&	communication	&	gpu cluster	&	No \\
\hline
ASTRA Toolbox	&	tomography	&	 \specialcell{MATLAB \\ Python}	&	compute	&	desktop/server	&	No \\
\hline
CISI	&	digital forensics	&	Java	&	compute	&	desktop	&	No \\
\hline
\specialcell{KM3NeT \\ L0-Trigger}	&	particle physics	&	Python	&	compute	&	gpu cluster	&	No \\
\hline
Parallel-Horus	&	computer vision	&	C	&	compute	&	gpu cluster	&	No \\
\hline
POP	&	climate modeling	&	Fortran90	&	communication	&	supercomputer	&	No \\
\hline
SAGECAL	&	radio astronomy	&	\CC	&	compute	&	gpu cluster	&	Yes \\
\hline
\end{tabular}
    \caption{Overview of the GPU application case studies.}\label{tab:applications}
    \vspace{-0.5cm}
    \end{table}

{\bf Parallel Ocean Program} (POP)~\cite{pop} is an ocean general circulation model. POP is a large Fortran 90 code that has been in 
development and use for a long time. While the code does not contain any particular computational hotspots, we have ported the 
equation of state and vertical mixing computations to the GPU. The main challenges were overlapping computation with CPU-GPU 
communication and host language integration. The kernels have a relatively low arithmetic intensity and not much opportunity for code 
optimizations.

{\bf Parallel-Horus}~\cite{parallelhorus} is an image processing library that automatically parallelizes image algebra operations. 
One of the more challenging aspects in integrating GPU kernels was to deal with the separate GPU memory. 
For most applications that use Parallel-Horus, 2D convolution 
is the most time consuming operation, for which we have implemented a highly-optimized and auto-tuned GPU kernel~\cite{convolution}.

{\bf SAGECal}~\cite{sagecal} is a radio interferometric calibration package supporting a wide range of source models. SAGECal is 
intended to run on various platforms from GPU clusters to low-end energy-efficient GPUs. We have significantly improved the 
performance of the original GPU code by removing the use of dynamic parallelism and changing how the problem is mapped to threads and 
thread blocks. Using intrinsics to accelerate the computation of trigonometric functions also significantly improved 
performance.

\section{GPU Programming Challenges}~\label{sec:gpuprogramming}

This section lists the challenges encountered, and the lessons learned, while building the applications described in 
Section~\ref{sec:applications}. Many of these originate from the fact that the GPU application is based 
on existing software.

\subsection{Dealing with Separate Device Memory}

The fact that GPUs have separate device memory introduces much complexity.
First of all, one needs to decide how to manage the GPU memory itself. Allocating and freeing memory can be costly operations and as such memory 
allocations are preferably reused over time. 
Furthermore, the GPU programmer needs to consider what policy to implement for when device 
memory is full, or when it can be more efficient to forgo using device memory and stream data directly from host memory instead.
There are two main perspectives to consider here, depending on whether a library or an application is being developed.

From a library perspective, the application using the library either manages device memory or trusts the library to handle it. In case the library 
acts as a drop-in replacement for an existing non-GPU library, the library needs to manage device memory and automatically transfer data. 
For example, in the Parallel-Horus image processing library it is not known in what order the operations will be called and
when data needs to be present in host or device memory. In this case, we extended the state-machine used inside the library to keep track of 
distributed memory to also keep track of host or device copies of the data. This allows the library to insert memory 
transfers only when necessary and as such many, otherwise redundant, data transfers can be eliminated.
%

When developing a GPU application, 
the programmer has a complete view of what needs to be executed on the GPU and in what order. In 
particular for applications that require low latency GPU operations, it can be important for performance to overlap the data 
transfers with computations on the GPU. We have used performance modeling to estimate the performance impact of various overlapping 
techniques~\cite{performancemodel}.

For applications that require frequent, asynchronous, or high-throughput data transfers between host and device memory, it may be necessary to change the way in 
which host memory is allocated, as these allocations will need to be pinned and page-aligned.
This could require extensive changes to the original application and may prove difficult for applications in object-oriented languages
such as Java, MATLAB, or modern Fortran.

\subsection{Host Language Integration}

In many cases, only a relatively small part of an existing application is ported to the GPU, and as such a large part remains in the original programming 
language.
In general, there are two ways of integrating GPU code into applications that are not written 
in C/\CC.  Either you write the host code in C/\CC{} and use some foreign function interface to call that 
host function from the original application, or you use the language bindings that are available for the 
OpenCL or CUDA driver APIs in the original program's language.
We now briefly discuss how GPU code might be integrated in several different languages that we have used.

{\bf \CC{}}. 
While OpenCL and CUDA have been designed for C and \CC{}, their runtime APIs were initially not object oriented. 
OpenCL now has an official \CC{} header, but for CUDA a high-level object-oriented runtime API is still lacking. 
As such, even for \CC{} there are separately developed API wrappers to create a modern \CC{} API for CUDA programming~\cite{cudaapiwrapper}.

{\bf MATLAB} offers three different ways to execute GPU code into applications. Firstly, through arrays that can be declared using a GPU-enabled array type.
Point-wise arithmetic operations on the array are lazily evaluated and compiled into CUDA kernels on first use. 
Secondly, PTX-compiled CUDA kernels can be loaded as functions into MATLAB.
Finally, the MEX interface allows to call C/\CC{} functions from MATLAB, which could in turn call GPU kernels.

If you have complex kernels that do more than point-wise operations, you are limited to option 2 and 3. If you also need fine-grained control over GPU memory, 
for example to reuse GPU memory allocations across kernel invocations, or to control exactly when data is transferred between host and device memory, you have to 
write the host code in C/\CC{} and use the MEX interface. 

{\bf Fortran} has fewer options. There is CudaFortran offered by the PGI compiler, which requires the GPU kernels to be written as CudaFortran subroutines. 
When working on the Parallel Ocean Program, we have written C functions around our kernels and some of the CUDA runtime API functions
to interface the kernels written as CUDA/C code from Fortran 90.

{\bf Java} has a number of different wrappers available for integrating GPU code. The direct language bindings for OpenCL and CUDA, JOCL and JCuda, are most 
commonly used, but object-oriented programming that adheres to common Java conventions requires additional wrappers. As most Java GPU libraries rely on open 
source contributions, it is often hard for these projects to keep up with the latest features.

{\bf Python} has a great number of options, ranging from array, data frame, or tensor libraries with GPU support to language bindings for OpenCL and CUDA
through PyOpenCL and PyCuda. The latter are not direct mappings of the C APIs and as such allow GPU programming in a way common to Python programmers.
However, both projects rely on open source contributions to be kept up to date. 
For example, initial basic \CC{}
support in the kernel language already existed at the time of CUDA 1.x, but PyCuda still requires kernels to have C linkage. 


\subsection{Optimizing Code}

One of the most time-consuming aspects of software engineering GPU applications is performance debugging and code optimization. 
With {\em code optimization} we mean the process of making changes in the kernel code with the aim to improve kernel performance.
Applying code optimization is often necessary to achieve high performance.


The roofline model~\cite{roofline} is an often used tool for analyzing whether the performance of a code is bandwidth or compute 
bound. While the original roofline model was not specifically developed for GPUs, it has proved an incredibly useful tool for GPU 
programming. Many adaptations of the roofline model have been introduced, for example a cache-aware roofline 
model~\cite{ilic2013cache}, a quantitative model for GPU performance estimation~\cite{konstantinidis2017quantitative}, or one
that includes PCIe data transfers~\cite{performancemodel}.
While performance models help to understand the bottlenecks in the code, it is still very hard to fully explain the performance at run time.

In general, it is important to understand that moving data around is much more expensive than computing on it. As such, many of the code optimizations that 
are applied to GPU code concern improving data access patterns, exploiting specific memories as caches, and reusing data. Code optimizations that focus specifically
on improving the computations, without changing data access patterns are in our experience quite rare.

In the end, the most frequently applied code optimization is to vary the amount of work per thread and thread block. This technique 
is referred to under many different names, including thread-block-merge~\cite{yang2010gpgpu}, thread 
coarsening~\cite{zoppetti2000automatic}, or 1xN Tiling~\cite{ryoo2008program}. The reason that this particular code optimization is 
so effective is because it improves upon many different aspects of the code, including reducing the number of redundant instructions 
across threads and thread blocks, improving the data access pattern to maximize locality, and improving the amount of reuse of data 
and/or intermediate computations.

However, in the end the GPU programmer is left with a large design space of how to parallelize the computation and map that onto threads, how many threads to use 
in each thread block dimension, how much work to give to each thread and thread block, and so on. These parameters are hard to predict and the optimal 
configuration can be very hard to find by hand~\cite{dedispersion,kerneltuner}. As such, we have made extensive use of auto-tuning techniques in many different 
applications to ensure optimal and portable performance of our applications across different GPUs and problem dimensions.

\section{Research Software Challenges}~\label{sec:gpurseing}

This section presents the challenges and lessons learned with regard to
the research software engineering aspects of developing GPU research software.

\subsection{Software Engineering Practices}

It is important to realize that GPU research software is often based on existing software developed by scientists with no formal training in software 
engineering~\cite{hettrick2014uk}. In addition, research software is often developed within projects of fixed duration with no resources for long-term 
maintenance~\cite{goble2014better}. The result is that research software often does not follow software engineering best practices, for example having no, 
little, or outdated documentation, no or only few automated tests, and code optimizations that might no longer be relevant.

\subsubsection{Testing}

Despite the large number of frameworks that seek to simplify GPU application development, a majority of GPU 
applications is still developed using low-level languages such as CUDA and OpenCL\@. This is in part due to 
the tension between increased abstraction levels and performance requirements that demand developer control 
over all aspects of the hardware. Another practical reason is that many tools and frameworks are developed 
as academic works, e.g.\ part of a PhD project, with no guarantee of sustained maintenance, making it a risk 
to use such a framework as a basis for a new project.

Unfortunately, not much CUDA and OpenCL code that is used in publications is tested. For a long time 
there were no testing frameworks that actively supported testing of GPU code. That is why we have extended 
Kernel Tuner~\cite{kerneltuner} to support testing GPU code from Python.

\subsubsection{Dealing with existing optimizations}



Several of the applications we have worked on included optimizations whose need had faded over time, as compilers improved or new hardware architectures 
were released. This phenomenon can also be observed in GPU applications.

%
For example, in the early days of CUDA programming increasing the work per thread required manually unrolling the 
inner-most loop, avoiding the use of arrays because of memory inefficiencies.
%
%
%
As it is often necessary to vary and tune the amount of work assigned to each thread, many 
applications contained multiple different versions of their GPU kernels with much code duplication 
as a result. Other developers used custom written code generators to avoid code duplication, resulting 
in code that is much harder for others to understand. Fortunately, with modern GPUs and 
compilers, it is possible to vary the work per thread dynamically.

Another example is from the Parallel Ocean Program that has been in development for a long time. A lot of 
the code we worked with was developed in the late nineties. Computer architectures were very different at 
the time, where compute was expensive and memory bandwidth was quite cheap. In modern processors, and in 
particular GPUs, this is quite the opposite, which means that some of the patterns in the original code 
could now be seen as unnecessary optimizations.


In general, it is very important to realize that the code that you develop may well outlive the hardware you 
are currently developing for. Trusting the compiler to handle the simple cases of code optimization
helps to improve the maintainability and portability of the code. As of CUDA 7.0 and OpenCL 2.x, \CC{}11 is 
supported in the kernel programming languages, improving readability and 
maintainability of GPU code. In addition, templates can be used to avoid code duplication.


\subsubsection{Selecting the right starting point}

Selecting the right starting point for developing GPU kernels is crucial. The first implementation for a GPU 
kernel is commonly called the {\em naive} version, which should be simple enough to allow the GPU to exploit 
massive data-parallelism. The name, however, suggests that there is something wrong with this code, which is 
definitely not the case. The naive code is a crucial part of the GPU application. Even if it is not executed 
in production, it should be stored along with the code, because it is necessary for 
understanding and testing the optimized and tunable versions of the code.

\subsection{Targeting problems at new scales}

An important question that is easy to overlook is whether the existing research software, on which the GPU code 
will be based, is actually capable of solving the problem at hand. This may sound obvious, but it is 
important to realize that moving to the GPU is often motivated by the desire to increase in scale and complexity.
The research group is developing GPU code because they want to address larger, more computationally demanding
problems at some larger scale or at a higher resolution.

At different scales it is often even required to use different algorithms.
Not just from a performance perspective, but also because the original code was developed to solve a problem
at a different scale, and might simply not produce meaningful results when executed on a larger problem.
Secondly, rounding errors and other numerical inaccuracies may accumulate in a stronger fashion when the
problem size changes.

For example, ocean models at coarse resolutions often use parameterizations to represent subgrid physical processes that are too small to be 
resolved at the current scale. However, at higher resolutions, we can do without these parameterizations because the 
parameterized physical processes are already resolved. As such, using the same algorithm without
modification at both resolutions would not produce correct results~\cite{gent2011hindsight}.

We have also observed this when working on the L0-trigger pipeline for the KM3NeT neutrino telescope, where the algorithms 
that were in place operated on the pre-filtered L1 hits.
However, using these algorithms to correlate L0 hits led to nearly all hits 
being correlated, as such the trigger pipeline did not work at all for the unfiltered L0 hits. 
In the end, we had to develop entirely new correlation criteria and community detection algorithms for the GPU pipeline.


\subsection{Software Sustainability}

Lago et al.~\cite{lago2015framing} define {\em software sustainability} as ``the capacity to endure and preserve the function of a 
system over an extended period of time''. In contrast to narrower definitions that only consider software sustainability as a 
composite measure of software quality attributes~\cite{venters2014blind}, the definition by Lago et al. also captures social aspects 
that are crucial for research software projects to endure.

Translating and parallelizing existing research software for the GPU during a short-lived collaboration creates a risk with regard to 
software sustainability. Programming GPU applications remains a specialized field that requires advanced technical knowledge and 
specialized programming languages. The GPU applications that we have developed were often part of temporary collaborations between 
different research groups. Once the project is finished, who is going to maintain the newly developed GPU code?

Without significant changes to the way in which scientific projects and software development for research is funded, there is no obvious solution that we are 
aware of. We can only recommend to think about software sustainability from the start. Measures that could be taken are involving the original developers of the 
code in the GPU development process, as well as documenting why and how the GPU code differs from the existing code.

However, involving the original developers in the development of a GPU version is not always possible. For example, in 
one project we could not get security clearance to access the internal version control servers, forcing
the two groups to develop on separate repositories.
%
%
Another issue that can occur is that some developers take offense when someone else starts to 
make significant changes to their code. This could be because accelerating their code can be interpreted as a sign 
that the original developers did something wrong, or simply because the original code had been extensively validated and any significant changes 
to the code may require to redo the validation.

Another take on the sustainability problem is to use a domain-specific language (DSL) to introduce a separation of
concerns between what needs to be computed and how the computations are mapped to the target hardware platform. This approach is, for example,
currently being researched in the weather and climate modeling community in Europe~\cite{lawrence2018crossing}. 

\subsection{Evaluating results}

Evaluating the output quality is one of the main challenges in developing GPU research software. When code is ported to the GPU, the output will not be 
bit-for-bit the same as it was on the CPU. There are many reasons for this.

Unlike the first programmable GPU architectures, modern GPUs fully implement the IEEE floating-point standard. Other factors do, however, introduce 
differences in the results. Floating-point arithmetic is not associative and as parallelization changes the order of operations, the results from a parallel 
algorithm will inherently be different from the sequential version, and will depend on the exact parallelization strategy and parameters chosen. Other 
reasons include the availability of different instructions (such as fused multiply-add instructions), 
the use of extended precision (80-bit floating point) in many CPUs, and different compilers that optimize differently.

RSEs of GPU applications will have to live with the fact that the results are not bit-for-bit the same. Whether 
that is a problem depends on the application. To analyze whether these differences matter for the application 
requires that you understand what is being computed and why, which requires close collaboration with the scientists 
that developed the original application.

It can also be difficult to determine whether the differences in results are due to a bug or due to the difference 
in parallelization and compilers. Testing can help to increase confidence in the GPU application, but we have
learned the hard way that testing with randomly generated inputs can give a false sense of correctness.
As such, we recommend that the test input is realistic enough to produce realistic output and to not only compare the GPU results with results obtained
on the CPU, but also assert that both outputs themselves are correct.

\subsection{Evaluating application performance}

Setting up experiments for performance comparisons of GPU applications,
and in particular presenting the performance improvement
over earlier versions, can be quite complicated.


In the early days of GPU Computing, many papers reported spectacular performance results. With 
their paper ``Debunking the 100X GPU vs. CPU myth''~\cite{lee2010debunking}, scientists from Intel made it 
clear that many of these performance results were based on unfair comparisons. In many cases, 
the performance of a highly-optimized GPU kernel was compared against an unoptimized, sometimes not even 
parallelized CPU kernel. Lee et al.~\cite{lee2010debunking} demonstrated for a large range of benchmark 
applications that if both the GPU and the CPU implementations are fully optimized, the performance difference 
between the two is usually within the range of the theoretical performance difference between the two platforms.

In computer science, the discussion around presenting performance results often assumes that the 
only question worth answering is: ``Which of these two processors is the most efficient for algorithm X?''. On 
the other hand, RSEs often work with scientists from different fields,
and thus receive questions such as ``How much faster is that MATLAB/Python code 
that I gave you on the GPU?''. This puts the RSEs in a difficult position. You would like to be able to tell the 
broader scientific community the fact that the application X now runs Y times faster.
However, that result on its own is currently very hard to publish, unless you are also willing 
to spend several more months to also optimize the CPU code and make a `fair' performance comparison between the two
processors. 

Not being able to publish about your work could ultimately limit career advancement, because while RSEs are tasked
with creating research software, they are often judged by metrics such as scientific publications~\cite{rse-state-of-nation}.



When it comes to presenting performance comparisons, it should be absolutely clear what is being 
compared against what, what hardware is being used by each application, and why that comparison is a realistic and representative use case. 
It is important that the experimental methodology is sound and is documented so that others may reproduce the results.

In their paper ``Where is the data?'', Gregg and Hazelwood~\cite{gregg2011data} point out that it is crucial 
to understand where the input and output data of a GPU kernel is stored. This strongly depends on the 
rest of the application. It could be that the GPU kernel is part of a pipeline of 
GPU kernels and that it is safe to assume that the input and output data are present in device memory.
However when this is not the case, the data transfer between host and device should be included in performance comparisons.

It is also possible to normalize results in different ways, for example based on hardware purchase costs, energy consumption, theoretical peak performance, 
person-months spent on code optimization. It all depends on the application whether such normalizations have any relevance, and they do not solve the problem in 
general. In addition, these normalizations can make it harder to interpret and compare the presented performance results across different publications.

\section{Conclusions}\label{sec:conclusions}

GPUs are a very attractive computational platform because they offer high performance and energy efficiency at relatively low cost. 
However, developing GPU applications can be challenging, in particular in a research software engineering context. 
For each of the case studies included in this paper, which span a wide range of research domains, programming languages, and target hardware platforms,
we have listed what the main challenges were in developing the GPU application.
We have grouped the recurring challenges and the lessons learned into two categories, related to either GPU programming in general, or 
specifically to developing GPU applications as a research software engineer.

In summary, developing GPU applications in general requires fundamental design decisions on how to deal with separate memory spaces, integrate different 
programming languages, and how to apply code optimizations and auto-tuning.

Developing GPU research software comes with a number of specific challenges, with regard to software engineering best practices, and the quantitative and qualitative
evaluation of output results. In general, we recommend to carefully select and if needed rewrite the original application to ensure the starting point is of
sufficient code quality and is capable of solving the problem at the scale the GPU application is targeting.
When performance comparisons of different applications are of interest to the broader scientific community it is important that RSEs can publish those results, 
both for the community to take notice of this result and for the RSEs to advance in their academic career. Finally, we note that software sustainability remains 
an open challenge for GPU research software when RSEs are only involved in the project on a temporary basis.

\bibliographystyle{splncs04}
\bibliography{paper}

\end{document}